# Co-doped $La_{0.5}Sr_{0.5}TiO_{3-\delta}$ : A Diluted Magnetic Oxide System with High Curie Temperature

Y. G. Zhao[¶], S. R. Shinde, S. B. Ogale[*], J. Higgins, R. J. Choudhary, V.N. Kulkarni[§], R. L. Greene, T. Venkatesan.

Center for Superconductivity Research, Department of Physics, University of Maryland, College Park, MD 20742-4111.

S. E. Lofland, C. Lanci

Department of Chemistry and Physics, Rowan University, Glassboro, N.J. 08028-1701.

J. P. Buban, N. D. Browning

Department of Physics, University of Illinois at Chicago, 845 West Taylor Street, Chicago, IL 60607-7059.

S. Das Sarma

 Condensed Matter Theory Center, Department of Physics, University of Maryland, College Park, MD 20742-4111.

A. J. Millis

Department of Physics, Columbia University, 538 West 120[th] Street, New York, New York 10027.

Abstract

Ferromagnetism is observed at and above room temperature in pulsed laser deposited epitaxial films of Co-doped Ti-based oxide perovskite ($La_{0.5}Sr_{0.5}TiO_{3-\delta}$). The system has the characteristics of an intrinsic diluted magnetic semiconductor (metal) at low concentrations (< ~ 2 %), but develops inhomogeneity at higher cobalt concentrations. The films range from being opaque metallic to transparent semiconducting depending on the oxygen pressure during growth and are yet ferromagnetic.

Induction of ferromagnetism in non-magnetic host materials by doping of magnetic impurities in dilute concentrations has gained considerable attention recently because of the addition of spin functionality to the original property-set of the concerned host [1], and thereby its potential applications in spintronics [1,2]. Considerable success has been achieved in the synthesis of diluted magnetic semiconductors (DMS) based on III-V semiconductors [3]. It has been shown that the ferromagnetism in DMS depends strongly and non-trivially on the carrier density and the concentration of the magnetic ions [4], as well as on the magnetic coupling between the carriers and the magnetic ions.

More recently room temperature ferromagnetism has also been reported in some magnetic impurity doped oxide based systems [5-8]. In this context the report of room temperature ferromagnetism in Co doped anatase $TiO_2$ by Matsumoto et al. [8] is particularly interesting, opening an avenue for exploration of other titanium oxides. Unfortunately, attempts to induce ferromagnetism in Ti-based oxides such as $LaTiO_3$ and $SrTiO_3$ have not met with success [9]. Here we examine the case of cobalt doping in the mixed state of the end compounds $LaTiO_3$ and $SrTiO_3$, namely the Ti-based perovskite $La_{0.5}Sr_{0.5}TiO_3$, and show that ferromagnetism at and above room temperature can indeed occur in this system. The choice of the mixed state compound was guided by the fact that its electrical properties (carrier dynamics) change dramatically with Sr concentration, changing it from an insulator to a metal [10]. $LaTiO_3$ (x=0), shows nonmetallic behavior due to strong correlation [11], which also leads to antiferromagnetic ordering [11,12]. With slight La/O off-stoichiometry or Sr doping, the antiferromagnetic ordering disappears and an insulator-metal transition occurs [11-13]. The carrier density in $La_{0.5}Sr_{0.5}TiO_3$ is about $8\times10^{21}/cm^3$ [10].

In our work, thin films of $La_{0.5}Sr_{0.5}Ti_{1-y}Co_yO_{3-\delta}$ ($0 \leq y \leq 0.07$) were grown on $LaAlO_3$ (001) substrates at 700 °C by pulsed laser deposition [14] with KrF excimer laser pulses (248 nm). The targets were prepared by a standard solid-state reaction method. The energy density and repetition rate were 1.8 J/cm$^2$ and 10 Hz, respectively. Depositions were performed at different oxygen pressures over the range from 3 x 10$^{-6}$ Torr to 10$^{-2}$ Torr. After deposition, the films were cooled to room temperature at 30 °C/min at the same pressure, and were characterized by various techniques.

Figure 1(a) shows the x-ray diffraction pattern of $La_{0.5}Sr_{0.5}Ti_{0.985}Co_{0.015}O_{3-\delta}$, indicating a single-phase perovskite structure with (00l) orientation. Similar data were also obtained for films with other concentrations and deposition pressures. The rocking curve full width at half maximum (FWHM) is ~0.3°, establishing the high orientational quality of the films. In Fig. 1(b) we compare the x-ray diffraction patterns (002 planes) for $La_{0.5}Sr_{0.5}Ti_{1-y}Co_yO_{3-\delta}$ films with y = 0.0 (undoped) and y = 0.015. A small shift of the film (002) peak to lower 2θ value (corresponding to expanded $d_{002}$) can be clearly noted; while the reference (002) $LaAlO_3$ substrate peak for the two cases overlaps exactly. This indicates incorporation of cobalt atoms into the matrix. Fig. 1 (c) shows the magnetization (M) for the $La_{0.5}Sr_{0.5}Ti_{0.985}Co_{0.015}O_{3-\delta}$ film grown at an oxygen pressure of 10$^{-4}$ Torr measured by SQUID (5-300K) and VSM (>300 K). It is clear that the film exhibits high temperature ferromagnetism with a Curie temperature close to 450K. Indeed, the inset shows the hysteresis recorded at 423 K, showing a well-defined loop with a coercivity of ~150 Oe. The nature of M-T curve is rather peculiar and its origin (whether related to the specific nature of the carrier-induced mechanism or materials related) is not understood at this time. Another inset in Fig. 1(c) shows the scanning transmission electron microscopy (STEM, JEOL, 0.14 nm resolution) cross-section image of $La_{0.5}Sr_{0.5}Ti_{0.985}Co_{0.015}O_{3-\delta}$ film, which confirms the high crystalline quality of the film. Similar STEM image quality was also seen for all other samples. However, in samples with y=0.07 EELS data showed inhomogeneity in cobalt concentration with suggestion of dopant clustering over 1-2 nm size scale. The results discussed below are therefore limited to films with low cobalt concentration of y=0.015, for which no such inhomogeneity was detected.

Figures 2(a,b,c) show the temperature dependence of resistivity ($\rho$) for three cases of single-phase perovskite films of $La_{0.5}Sr_{0.5}Ti_{0.985}Co_{0.015}O_{3-\delta}$, grown at the oxygen pressures of $10^{-2}$, $10^{-4}$, and $3 \times 10^{-6}$ Torr, respectively. The typical form in all cases is weak metallicity at and below room temperature and semiconducting nature at lower temperature. The temperature at which the metal-semiconductor transition occurs is seen to shift to higher temperature with increased oxygen pressure during growth. The dependence of resistivity on $T^2$ was noted to exhibit good linearity over the region of metallicity, suggesting a strong electron-electron interaction in this compound.

Fig. 2(d) shows the 5K and 300K resistivity values, which follow similar growth pressure dependence and span almost three decades on the resistivity scale. The resistivity decrease with decrease in oxygen growth pressure implies that conduction in films is controlled significantly by oxygen vacancy concentration. It may further be noted from Fig. 2(d) that the out of plane lattice parameter also shows a relaxation by up to 1.5% over the same range of growth pressures, and the direction of change appears consistent with the expected change in the oxygen vacancy concentration.

In Fig. 2(e) we show the dependence of magnetization on resistivity at 5K for a set of $La_{0.5}Sr_{0.5}Ti_{0.985}Co_{0.015}O_{3-\delta}$ films grown at different pressures. The first, third and fifth filled black circles from right to left correspond to the data of Fig. 2 (a), (b) and (c), respectively. We also show the carrier density estimated from Hall data on the right y-axis for a few cases of interest. It may be noted that ferromagnetism with a significant moment is observed only in samples in the intermediate resistivity range of about $10^2$ to $10^4$ μΩ-cm, and the dependence is non-monotonic with an asymmetric bell shaped nature. Interestingly however, over the same range the carrier density changes monotonically. The moment M near the peak of the bell shaped curve is significantly higher than its value of ~1.72 $μ_B$/Co for pure cobalt metal. The large value of M and its dependence on growth pressure may imply presence and evolution of a high-spin and low spin admixture of Co(II) and Co(III) ions, conditions known to occur for octahedrally coordinated cobalt as in perovskite matrices [15,16]. Other possibilities such as a partial transferred moment on Ti may also have to be addressed by further experiments.

While these data strongly suggest some role of carrier density ($n_{carrier}$) and dynamics in the occurrence of ferromagnetism, it is difficult to offer a simple physical picture at this stage due to the coexisting contributions of cobalt ions and oxygen defects to the transport. It may be seen from Fig. 2 (e) that the data in our case fall under the regime of cobalt concentration $n_{Co} \ll n_{carrier}$. According to standard DMFT calculations [4], the regime of $n_{Co} \ll n_{carrier}$, which is complementary to the GaMnAs DMS regime, should not be a ferromagnet. In presence of strong disorder this regime should really be a spin glass (or other magnetic glass). Often, a large fraction of the local moments is just missing from the net moment for poorly understood materials reasons. In metallic

GaMnAs it is attributed to interstitial Mn, which couples antiferromagnetically to substitutional ones (also acting as compensating impurities) lowering net moments. Various annealing procedures are being followed for GaMnAs to suppress this Mn interstitial formation. If, in our case, some Co moments are missing at both high and low resistivity samples for reasons we do not understand (i.e. materials reasons), that could be an explanation for the bell shaped nature of the curve. In large resistivity samples percolation effects may be playing a role. If that is happening then, according to the theory of Kaminski and Das Sarma [17], magnetization may not saturate even down to 5K, leading to an apparent low value of M in the low carrier density samples. In the high carrier density, low resistivity sample, a competition with some possible glassy phase may effectively lower M. Glassy magnetic phases are more likely at higher carrier densities and/or lower Co densities. Unfortunately, at this exploratory phase of research it is hard to offer further insights into the mechanism, without sorting out potentially materials related issues.

In Fig. 2(f) are shown representative magnetoresistance data for the $La_{0.5}Sr_{0.5}Ti_{0.985}Co_{0.015}O_{3-\delta}$ film grown at $10^{-4}$ Torr. Significant negative magnetoresistance (MR = $(\rho_H - \rho_o)/\rho_o$ x 100 %) is seen only at very low temperatures when the resistivity was confirmed to exhibit variable range hopping (VRH) behavior. Such interesting behavior has also been reported in other VRH systems, and is attributed to an interplay of quantum interference effects at low fields and orbital shrinkage at high fields [18,19].

Finally, in Fig. 3 we show the magnetic and optical properties of the relatively resistive $La_{0.5}Sr_{0.5}Ti_{0.985}Co_{0.015}O_{3-\delta}$ film grown at $10^{-2}$ Torr. The optical transmission data and the picture in the inset show that film grown under such high pressure qualifies to be a transparent diluted magnetic system. The temperature dependence of M (inset) shows a $T_C$ close to ~550K, about 100K higher than that for the relatively conducting film grown at $10^{-4}$ Torr (Fig. 1(c)). Notably, the decrease of magnetization to zero value is gradual for the resistive film (Fig. 3, inset) as compared to that for the conducting film (Fig. 1(c)). Also, there is a small bump in the curve near the transition, which could be due to a small contribution of a secondary magnetic phase. In the case of more resistive samples a mechanism of magnetic polaron percolation has been suggested for DMS systems [17]. In such a scenario a more gradual build up of magnetization with lowering of temperature could be expected.

In conclusion, ferromagnetism is observed at and above room temperature in epitaxial $La_{0.5}Sr_{0.5}Ti_{1-y}Co_yO_{3-\delta}$ perovskite films deposited over a range of oxygen pressures. It is shown that ferromagnetism is realized in both opaque metallic as well as transparent semiconducting films. While, the dependence of the magnetic and transport properties on y and $\delta$ suggests a strong role of carrier density and dynamics, materials related issues can not be completely ruled out. At this stage it is not absolutely clear whether ferromagnetism in Co doped oxides, including LSTO, (or for that matter in many of the Mn doped III-V semiconductors, e.g. GaMnN, GaMnP, except perhaps GaMnAs) is really an intrinsic carrier induced ordering of the local moments, or an extrinsic effect of nanoclustering (of dopant or related compound) or some combination of both. This obviously needs to be further explored in subsequent work by us and by others.

This work was supported by NSF under the MRSEC grant DMR-00-80008, by DARPA (T.V, S.B.O) by US-ONR (S.D.S) and DARPA (S.D.S), and New Jersey Commission on Higher Education (S.L.). Two of the authors (J.P.B and N.D.B) would like to acknowledge DOE funding under grant DE-FG02-96ER45610 and NSF support for the purchase of the JEOL microscope under grant DMR-9601792.

¶   On leave from Tsinghua University, China.

*   Also at the Department of Materials Science and Nuclear Engineering, ogale@squid.umd.edu

§   On leave from Indian Institute of Technology, Kanpur, India.

Figure Captions:

Fig. 1: (a) A typical x-ray diffraction (XRD) pattern of the $La_{0.5}Sr_{0.5}Ti_{0.985}Co_{0.015}O_{3-\delta}$ film, (b) Comparison of XRD patterns for $La_{0.5}Sr_{0.5}Ti_{1-y}Co_yO_{3-\delta}$ films with y = 0 and 0.015, (c) M-T curve for the $La_{0.5}Sr_{0.5}Ti_{0.985}Co_{0.015}O_{3-\delta}$ film. The line represents five point averaged data. The inset shows hysteresis loop at 423 K. Second inset shows the STEM image.

Fig. 2: Temperature dependence of resistivity for the $La_{0.5}Sr_{0.5}Ti_{0.985}Co_{0.015}O_{3-\delta}$ films grown at (a) $10^{-2}$, (b) $10^{-4}$, and (c) $3 \times 10^{-6}$ Torr. The 300 and 5 K resistivity and the (002) lattice parameter measured at 300 K for the films grown at different oxygen pressures are shown in (d). The dependence of magnetization on resistivity at 5 K is plotted in (e). The carrier density is also shown for a few cases of interest. The magnetoresistance data are shown in (f).

Fig. 3: Optical transmission spectrum for the $La_{0.5}Sr_{0.5}Ti_{0.985}Co_{0.015}O_{3-\delta}$ film grown at $10^{-2}$ Torr. The transparency in the visible range is clearly demonstrated by the photograph at the upper left corner. The high temperature M-T curve for this film is shown in inset.

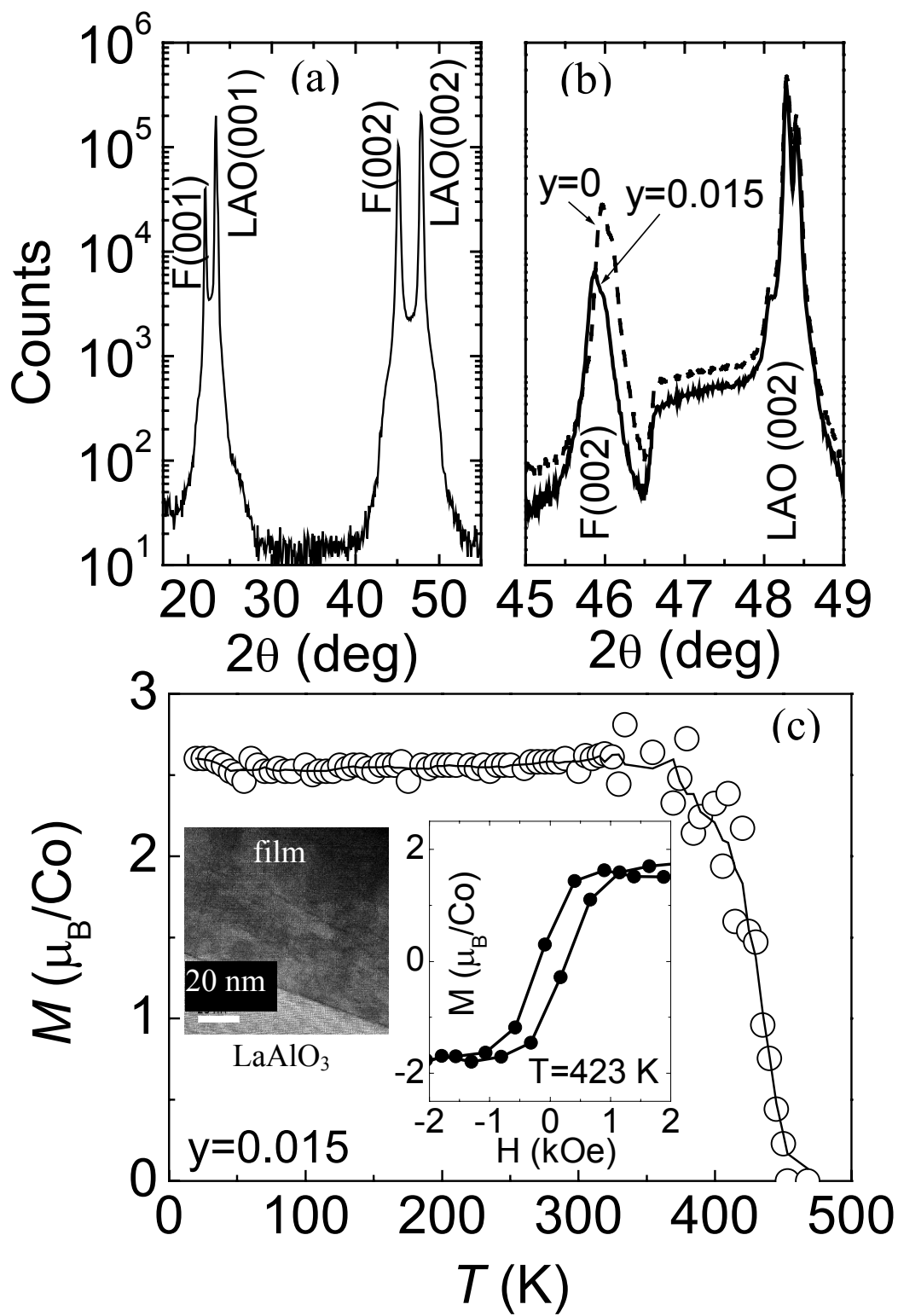

Figure 1, Zhao et al

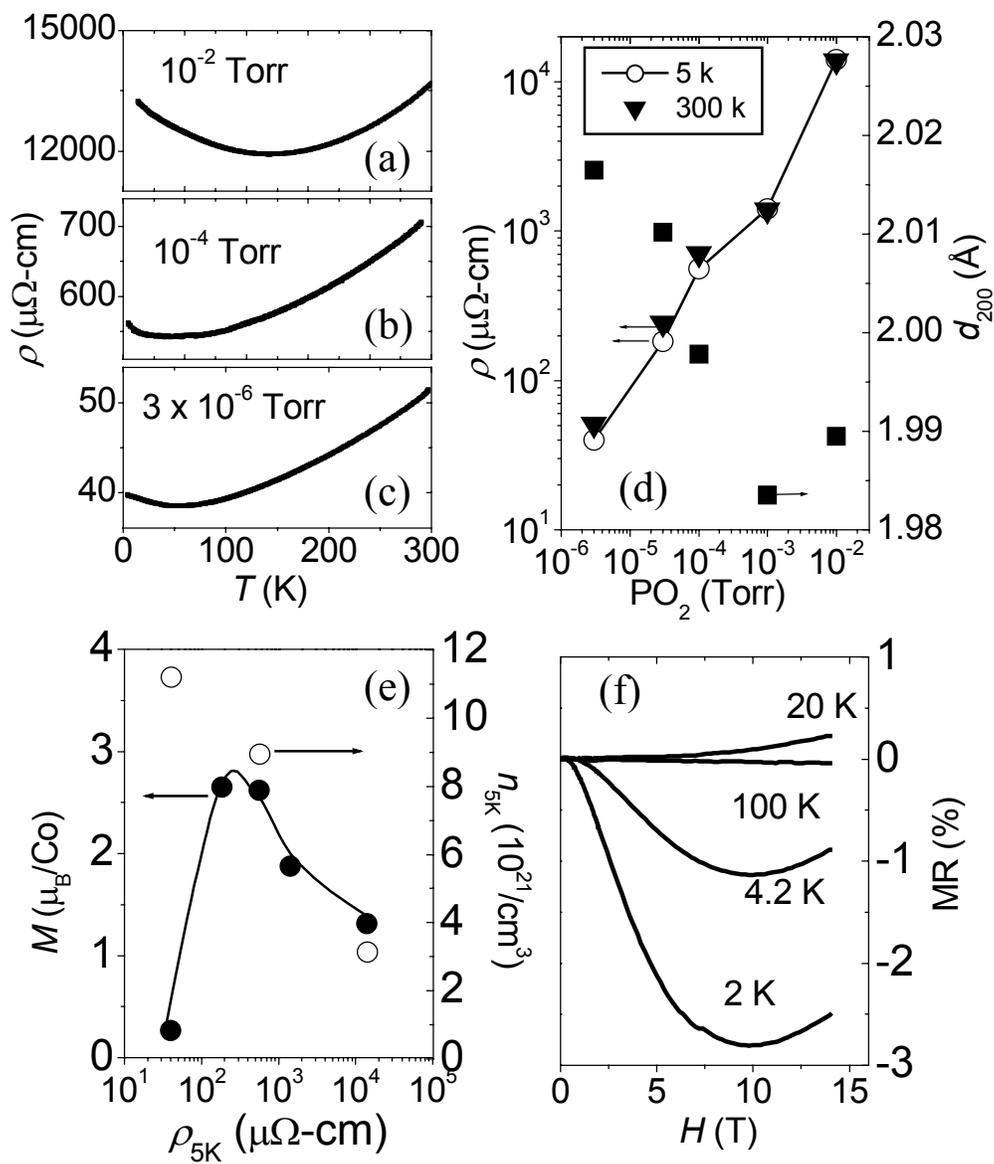

Figure 2, Zhao et al

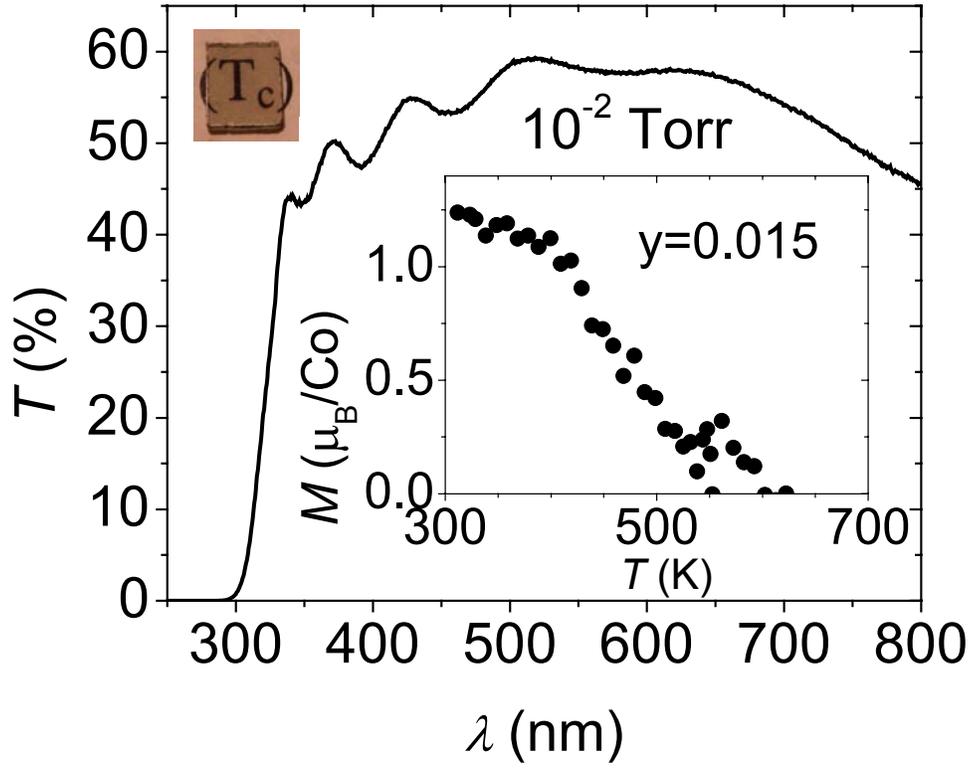

Figure 3, Zhao et al

15